\documentclass[
aip,
amsmath,amssymb,
preprint,%
]{revtex4}

\usepackage{graphicx}
\usepackage{dcolumn}
\usepackage{bm}

\usepackage[utf8]{inputenc}
\usepackage[T1]{fontenc}
\usepackage{mathptmx}
\usepackage{amsmath}
\usepackage{float}
\begin{document}


\title
{Adsorption anomalies in a 2D model of cluster-forming systems} 

\author{E. Bildanau}
\affiliation{ 
	Belarusian State Technological University, 13a Sverdlova Str., 220006 Minsk, Belarus
}%

\author{J. P\c{e}kalski}
\affiliation{ 
	Institute of Physical Chemistry, Polish Academy of Science, 01-224 Warsaw, Poland
}%
\affiliation{Department of Chemical and Biological Engineering, Princeton University, Princeton, New Jersey 08544, USA}

\author{V. Vikhrenko}%
\affiliation{ 
	Belarusian State Technological University, 13a Sverdlova Str., 220006 Minsk, Belarus}

\author{A. Ciach}%
\affiliation{ 
	Institute of Physical Chemistry, Polish Academy of Science, 01-224 Warsaw, Poland
}%

\date{\today}

\begin{abstract}
Adsorption on a boundary line confining a monolayer of particles self-assembling into clusters  is studied by MC simulations. We focus on a system of particles interacting via competing interaction potential in which effectively short-range attraction is followed by long-range repulsion, mimicking the so called SALR system. For the chemical potential values below the order-disorder phase transition the adsorption isotherms were shown to undergo non-standard behavior, i. e. the adsorption exhibits a maximum upon structural transition between structureless and disordered cluster fluid. In particular, we have found that the adsorption decreases for increasing chemical potential when (i) clusters dominate over monomers in the bulk, (ii) the density profile in the direction perpendicular to the confining line exhibits an oscillatory decay, (iii) the correlation function in the layer near the adsorption wall exhibits an oscillatory decay in the direction parallel to this wall. Our report indicates striking differences between simple and complex fluid adsorption processes.
\end{abstract}

\maketitle


\section{\label{sec:intro}Introduction}

Traditionally, adsorption is considered as a deposition of particles on a surface. In the case of simple atomic or molecular
species, polymers, liquid crystals, proteins, other biological objects it has received a huge amount of attention, 
considering different aspects of the phenomena. In particular, monolayer and multilayer adsorption including wetting, 
prewetting, different phase transitions and criticallity of adsorbed layers were investigated in detail~
\cite{roe:74:0, dash:75:0,kreuzer:86:0,kukushkin:98:0,jerome:91:0,patrykiejew:00:0,netz:03:0,bruch:07:0,rabe:11:0}. 
On the other hand, not much attention has been given to adsorption in monolayers on restricting walls.
If the adsorbed monolayer is modeled as a two-dimensional (2D) system, then the confining walls are one-dimensional (1D) 
and one can consider layers of particles formed in the neighborhood of the 1D walls as adsorbed layers. 
In the present study, we describe how particles interacting via nonmonotonical pair potential in which effectively
hard-core repulsion is followed by attraction well and repulsive tail (SALR system) adsorb on a straight 1D wall 
that confines a flat surface.

For 2D systems, the influence of confinement on pattern formation was previously investigated 
experimentally~\cite{antelmi:95:0,yu:06:0,yu:08:0,su:03:0,li:05:0,huang:06:0,haghgooie:06:0},
theoretically~\cite{tasinkevych:01:0,imperio:07:0, imperio:08:0,archer:08:0,chi:11:0}
and by computer simulations~\cite{almarza:16:0,pekalski:19:0}. When the pattern formation is induced by 
competing attractive and repulsive interactions the size and shape of the confinement was shown to be
crucial when the aim is to fabricate a defect free pattern~\cite{almarza:16:0} or a chiral structure~\cite{pekalski:19:0}.
The confinement effects on SALR clusters were described also in 1D case, where the aggregates were shown 
to induce spatial bistability~\cite{pekalski:14:1} or pressure decrease upon increase of the density~\cite{pekalski:14:0}.
To our knowledge, however, for particles interacting with isotropic competing interactions 2D adsoprtion isotherms have 
not been described so far.

In dilute SALR systems, formation of clusters has been observed when the particle density exceeds the value corresponding 
to the critical
cluster density~\cite{santos:17:0,litniewski:19:0}, analogous to the critical micelle density in surfactant mixtures.
While the ordered periodic phases in the SALR\cite{zhuang:16:0} systems have not been observed experimentally 
yet~\cite{royall:18:0,zhuang:16:2}, 
the cluster fluids, first reported in Ref.~\cite{stradner:04:0}, are quite often observed for various systems. 
Only recently, the effect of clustering on the adsorption on confining walls has been studied~\cite{litniewski:19:0}
and it was shown that the effect of clustering on the adsorption phenomena in 3D is very strong, and deserves further 
investigation.

In this work we study the entirely unexplored question of the effect of self-assembly into clusters on adsorption on a confining line in a 2D system of SALR particles.  
In the case of a single boundary line at $z=0$, the line excess amount or the Gibbs adsorption is defined as follows:
  \begin{equation}
  \Gamma(\mu^*) = \int_{0}^{\infty}(\rho (z) - \rho_b)dz
  \label{eq:gamma}
 \end{equation}
where $\rho(z)$ and $\rho_b$ are the average density at the distance $z$ from the wall and in the bulk, respectively, for fixed chemical potential $\mu^*$.

We calculate the adsorption isotherms for a triangular lattice model introduced earlier in Refs.\cite{pekalski:14:0,almarza:14:0} and summarized in sec. \ref{sec:model}. In the same section our Monte Carlo simulation method is briefly described. In addition, we calculate structural characteristics such as the cluster distribution in the bulk and near the wall, density profile in direction perpendicular to the wall and the correlation function in the layers parallel to the wall. The results are presented in sec.\ref{sec:results}, where the relation between the shape of the adsorption isotherm and the structure of the fluid is also discussed. We summarize our results and present our conclusions in sec.\ref{sec:concl}.

\section{\label{sec:model} The model and the simulation procedure}

In order to allow close-packed structure, we have used a triangular lattice with a lattice constant equal to the particle diameter. After Ref.\cite{almarza:16:0, almarza:14:0}, we assume the following interaction potential between the particles on the lattice sites:
\begin{equation}
	V(\Delta \mathbf{x}) = 
	\begin{cases}
		-J_1 \quad \textrm{for $|\Delta \mathbf{x}| = 1,$ \quad  (for nearest neighbors)} \\
		+J_2 \quad \textrm{for $|\Delta \mathbf{x}| = 2,$ \quad  (for third neighbors)} \\
		0 \qquad \textrm{otherwise} 	
	\end{cases} 
\end{equation}
where $-J_1$ and $J_2$ represent the energy of interparticle attraction and repulsion, respectively. We used the ratio $J_2/J_1 = 3$ as in Refs \cite{pekalski:14:0, almarza:16:0, almarza:14:0}. 

The thermodynamic Hamiltonian for our system has the following form:
\begin{equation}
	H = \frac{1}{2} \sum_{\mathbf{x}}\sum_{\mathbf{x'}}\hat{\rho}(\mathbf{x})V(\mathbf{x}-\mathbf{x'})\hat{\rho}(\mathbf{x'}) - \mu\sum_{\mathbf{x}}\hat{\rho}(\mathbf{x}) + U_w\sum_{\mathbf{x_0}}\hat{\rho}(\mathbf{x})
\end{equation}
where $\sum_{\mathbf{x}}$ is the sum over all lattice sites, $\sum_{\mathbf{x_0}}$ is the sum over the sites nearest to the wall (or walls),  $\hat{\rho}(\mathbf{x})$ is the occupation number. $\hat{\rho}(\mathbf{x})=1$ or 0 if the site with the coordinate $\mathbf{x}$ is occupied or vacant. $U_w$ is the interaction energy of a particle with the wall. It can be negative (attractive boundary), positive (repulsive boundary) or vanishing (neutral wall). In experimental systems, the long-range repulsion between the particles is often of electrostatic origin. In such a case, the walls interacting with the particles only at short distances are charge-neutral. In calculations, the dimensionless values $T^*=k_B T/J_1$, $\mu^*=\mu/J_1$, $J_2^*=J_2/J_1$, $h=U_w/J_1$ are used.

The phase diagram as well as the ground states of this system in the bulk were investigated in Ref.\cite{almarza:14:0,pekalski:14:0}.  As shown in Ref.\cite{almarza:16:0}  for the  stripe (lamellar) phase, 
the confinement between two parallel lines can drastically change the structure in the whole slit.
The presence of a single wall may change the particle arrangement in its vicinity, but the effect of a single confining line has not been studied in this model yet. 

In theory, the phenomenon of adsorption is considered as the deposition of particles on a planar boundary in a semi-infinite system, Eq.(\ref{eq:gamma}), but in the computer simulation, a stripe between two walls has to be modelled. We define the adsorption for our model by

\begin{equation}
 	\Gamma(\mu^*) 
 	\approx 1/2 \sum_{z=0}^{L-1}(\rho(z) - \rho_c)
 	\label{eq:gammal}
 \end{equation}
 where $\rho(z)$ and $\rho_c$ are the average density at the distance $z$ from the wall and in the central one third 
 part of the system, respectively, and both densities are calculated for the chemical potential $\mu^*$. For large 
 enough inter-wall distance $L$, $\rho_c$ should be the same as the density in the bulk, $\rho_b$; otherwise the 
 finite-size effects have to be taken into account.
The finite size effects can be studied by varying the wall-wall distance, but it goes beyond the scope of this work. 
In the direction parallel to the walls, periodic boundary conditions are specified. 

The Monte Carlo simulation procedure was carried out in the grand canonical ensemble according to the Metropolis 
algorithm with a standard importance sampling. 
We chose the distances between the walls $L$ and the system size along the walls $H$ the same as $L = H = 80$ to 
be sure that this distance is several times larger of the largest correlation length in the system. To verify that 
$\rho_c=\rho_b$, we computed $\rho_b$ in a system with periodic boundary conditions.

\section{\label{sec:results}Results }

\subsection{The ground State ($T^*=0$)}

As noted above, we assume that the interaction between the particles and the walls exists only in the rows closest to them.
The walls can be neutral $h=0$,  attractive $h<0$ or repulsive $h>0$. 

\begin{figure*}[htb!]
	\includegraphics{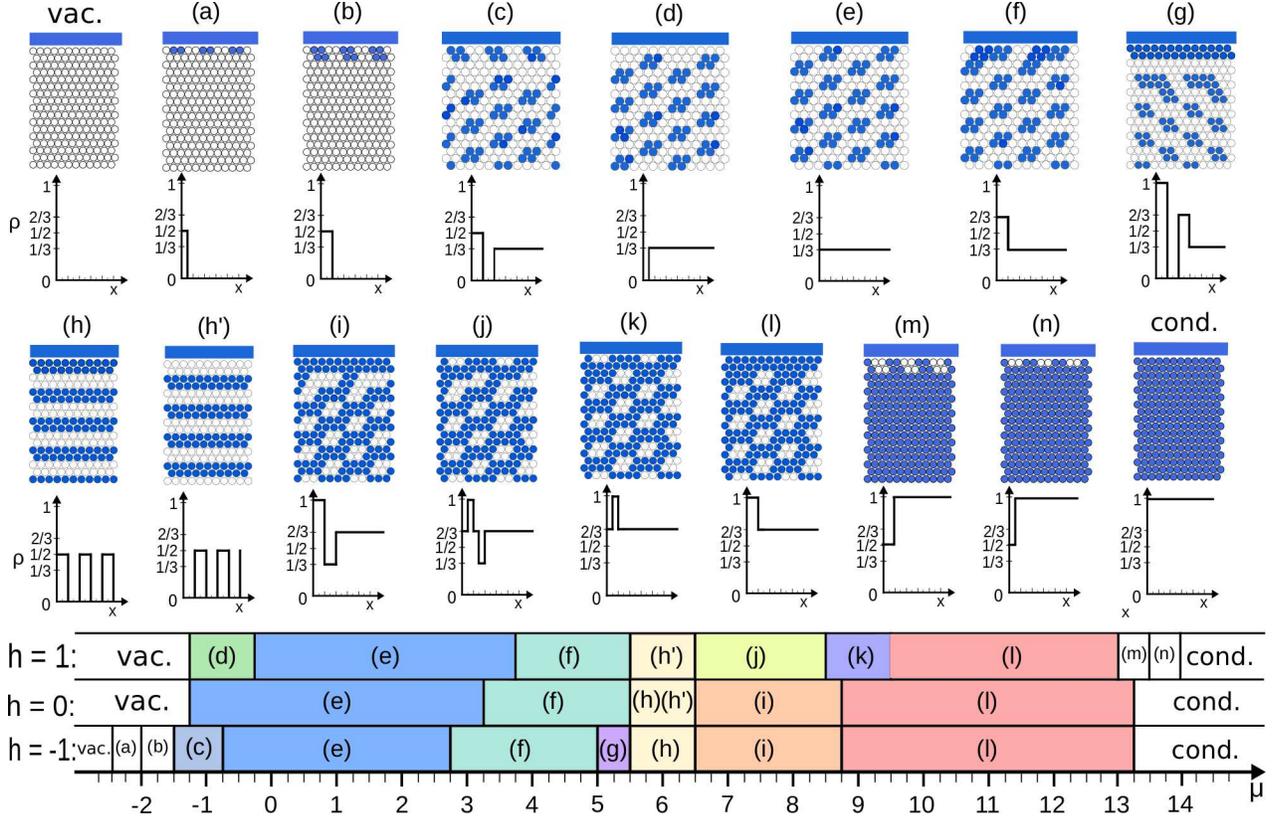}
	\caption{\label{fig:GS_ADS} The ground states ($T^* = 0$) for $h=-1,0,1$ and for different values of the chemical potential $\mu^*$. The snapshots present a region of the slit close to one of the confining walls. Blue and white circles represent occupied and empty sites, correspondingly. Below them the density profiles, $\rho(x)$, along the slit cross-section are presented. The lower panel presents regions of stability of the structures (a-n) in dependence of $\mu^*$ for the three different values of $h$. The regions denoted by vac. and cond. correspond to the vacuum and condensed phases, respectively.}
\end{figure*}

In the case of neutral walls ($h=0$), the ground states are cluster (e,f), lamellar (h) and bubble (i,l) 
phases (Fig. ~\ref{fig:GS_ADS}) that are similar to that in the bulk. The phase (e) only indicates absence of 
adsorption on the wall with homogeneous distribution of the average density in layers parallel to the wall. 
The lamellar phases (h) and (h') are strictly periodic in the direction perpendicular to the wall and filled
layers start directly from the attractive wall. The other phases are separated from the wall by different 
structures in two (f,l) or four (i) adjoining layers. For not neutral walls the same bulk structures appear
with more rich adsorbed structures within six adjoining rows at most.

It is interesting to consider an attractive wall and small values of the chemical potential corresponding to
the ordered cluster phase shown in panel (c) in Fig.\ref{fig:GS_ADS}. In the two rows closest to the boundary
(rows 1 and 2), the clusters are packed more densely than in the bulk. The two following rows (rows 3 and 4), 
however, are empty because of the repulsion between the third neighbors. As a result, the adsorption is negative,
$\Gamma=-1/3$. This counter-intuitive result showing that the attractive surface can lead to a desorption follows 
from the formation of the depletion (empty) zone following the adsorbed layer.
Let us focus on the vacuum that is the $T=0$ state of the disordered dilute phase studied at $T>0$ in the following sections.
A layer of thin or thick clusters is formed for
$-2.5<\mu^*<-2.0$ 
or
$-2.0<\mu^*<-1.5$ 
respectively. The adsorption is positive in the two cases, in contrast to the case discussed above. 
At 
$\mu^*=-2.0$ 
there is a discontinuous change from $\Gamma=1/2$ to $\Gamma=1$,  next at 
$\mu^*=-1.5$ 
the adsorption jumps from $\Gamma=1$ to $\Gamma=-1/3$, and at 
$\mu^*=-0.75$ from $\Gamma=-1/3$ to $\Gamma=0$.  

%

\subsection{The adsorption isotherms}
\begin{figure*}[htb!]
	\includegraphics[width=1\linewidth]{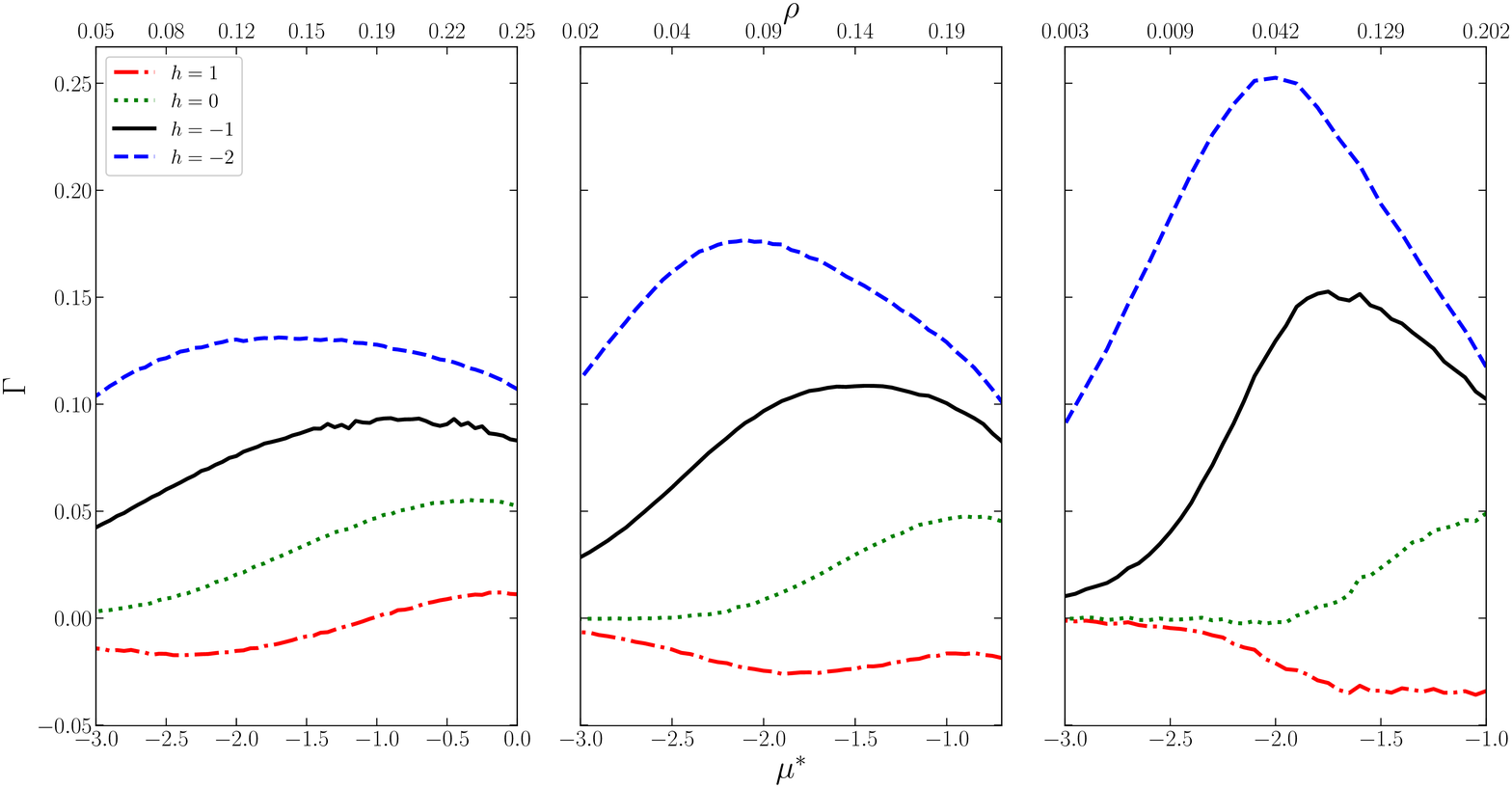} 
	\caption{\label{fig:ADS} Adsorption $\Gamma$ $versus$ the chemical potential $\mu^*$ for different values of the 
	wall-particle interaction
	 $h$ at the system temperatures $T^*=1.0$ left, $T^*=0.7$ central and $T^*=0.5$ right panels. $\Gamma$,
	 $T^*$ and $\mu^*$ are dimensionless.}
\end{figure*}

We investigated the behavior of the adsorption at various temperatures and the interaction with the walls (Fig. ~\ref{fig:ADS}).
The chemical potential values are restricted by $\mu^*_{pt} = -1.0, -0.7, 0.0$ for $T^* = 0.5, 0.7, 1.0$, respectively, 
to avoid the influence of phase transition effects on the adsorption phenomena.

For the walls with attracting character ($h<0$), the maximum of adsorption at the low temperature $T^*=0.5$ is 
observed at the value of the chemical potential, which corresponds to the disordered gas phase in the bulk.
At the lowest value of the chemical potential, the adsorption is weak due to very low density of particles. 
With increasing the chemical potential and correspondingly the bulk density, the adsorption increases as well 
up to an intermediate value of the chemical potential and then starts to decrease. 

In a warmer environment ($T^*=0.7, 1.0$), the adhesion of particles to the wall becomes less intense.
As a result, the maximum values of adsorption decrease with increasing temperature and the peak of the 
adsorption is smoothed out. It is of interest to investigate the effect of the formation of this peak 
in more detail. To this end, the partial adsorption is introduced as

\begin{equation}
    \Gamma(z) =  \sum_{x=0}^{z}(\rho(x) - \rho_c). 
    \label{eq:PartAds}
\end{equation}

\begin{figure}[htb!]
	\includegraphics[width=1\linewidth]{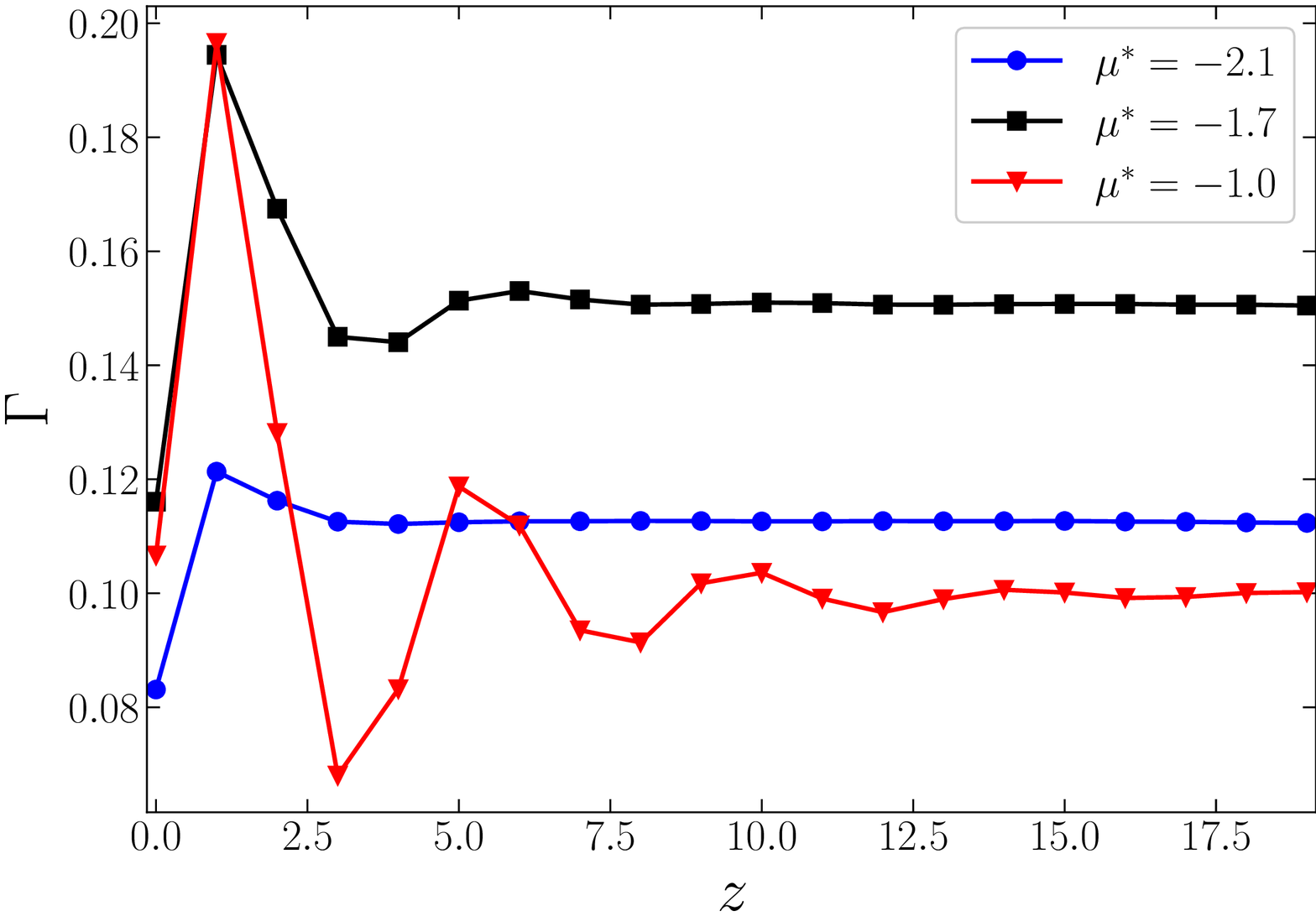}
	\caption{\label{fig:Par_ADS} The partial adsorption (\ref{eq:PartAds}) for the chemical potential 
	$\mu^*$ values smaller, at the adsorption maximum and above the latter at $T^* = 0.5$ and $h = -1$.} 
\end{figure}

The partial adsorption for the chemical potential values smaller, at the adsorption maximum and above the latter 
is shown in Fig. ~\ref{fig:Par_ADS}.

The partial adsorption is asymptotically approaching the total adsorption for the same chemical potential.
However, the deviations of the partial adsorption from the asymptotic value are growing when the chemical 
potential and bulk density are increasing. This is a consequence of the clusters formation in the system. 
The partial adsorption dependence on $z$ allows to estimate the distance from the wall where its influence spans. 
It strongly increases with the density and involve 4, 8 and 16 layers for $\mu^*=-2.1, -1.7$ and -1.0, consecutively.

\subsection{The density profile}

\begin{figure}[htb!]
	\includegraphics[width=0.7\linewidth]{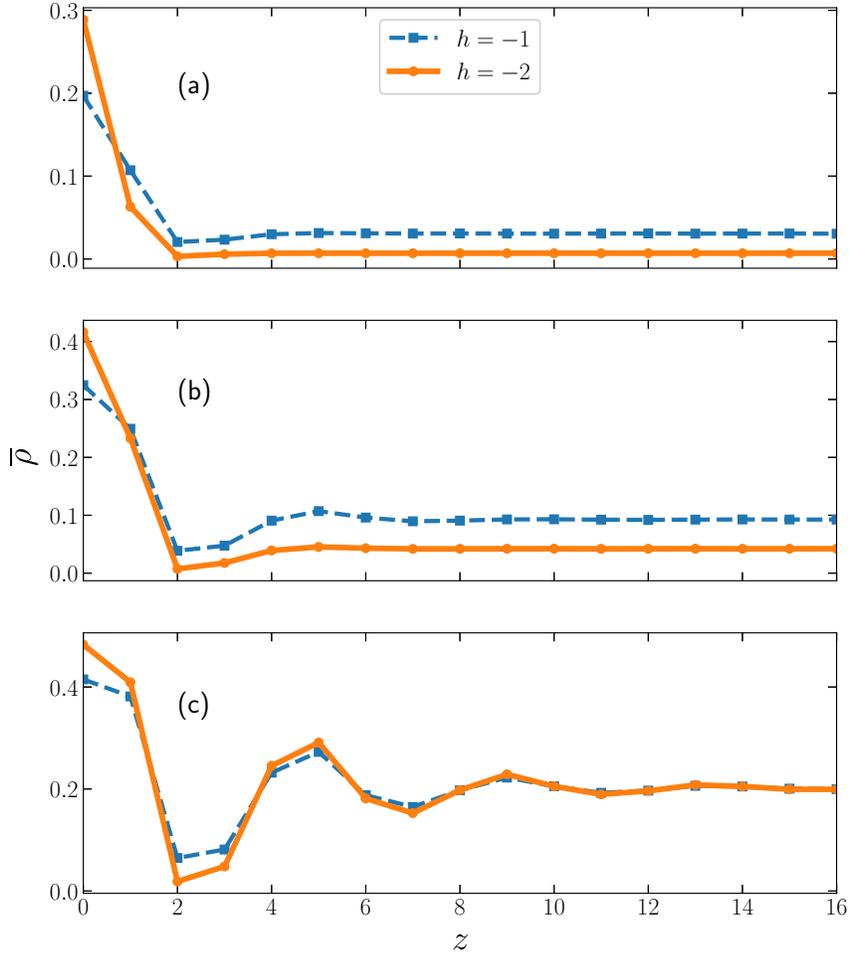}
	\caption{\label{fig:profiles} The density profiles in the near wall 
	region for different particle attraction to the wall, $h$, and for the temperature $T^*=0.5$. 
	There are 3 values of the chemical potential corresponding to the regions before, at and after
	the peak of the adsorption isotherm. The profiles are shown for $\mu = -2.6, -2.1$ (a), $\mu^*=-2.0,-1.7$ 
	(b), $\mu^* = -1.0, -1.0$ (c) for particle-wall interaction $h=-2$ (solid line) and $h=-1$ (dashed line). } 
\end{figure}

The distribution of particles in the near wall region demonstrates considerable changes with increasing the chemical
potential and correspondingly the bulk concentration (Fig. ~\ref{fig:profiles}). At low density at the system states
before and at the adsorption maximum four or eight nearest to the wall rows show deviation from the bulk density. 
The closest row is excessively populated due to attraction to the wall and the next row is populated as well because 
of the interparticle attraction of the first neighbours. Two subsequent rows are depleted in view of the repulsion
of the next-next-nearest neighbours. At low density of particles the influence of the wall decreases fast with the
distance to the wall. At larger concentration (for $\mu=-1$) the relay mechanisms transfer the density deviations 
for longer distances.

The qualitative behavior of the density profiles at higher reduced temperatures (0.7 and 0.9) remains the same. 
However, because of larger densities at $T^*=0.9$ the damped oscillations are visible already at the chemical
potential corresponding to the adsorption maximum. 

With an increase in the energy of interaction with the wall $h$, the density of adsorbed particles 
near it increases and the peak position of $\Gamma(\mu^*)$ shifts to the region of lower chemical potential
(Fig. ~\ref{fig:ADS}). 

\begin{figure}[htb!]
	\includegraphics[width=0.8\linewidth]{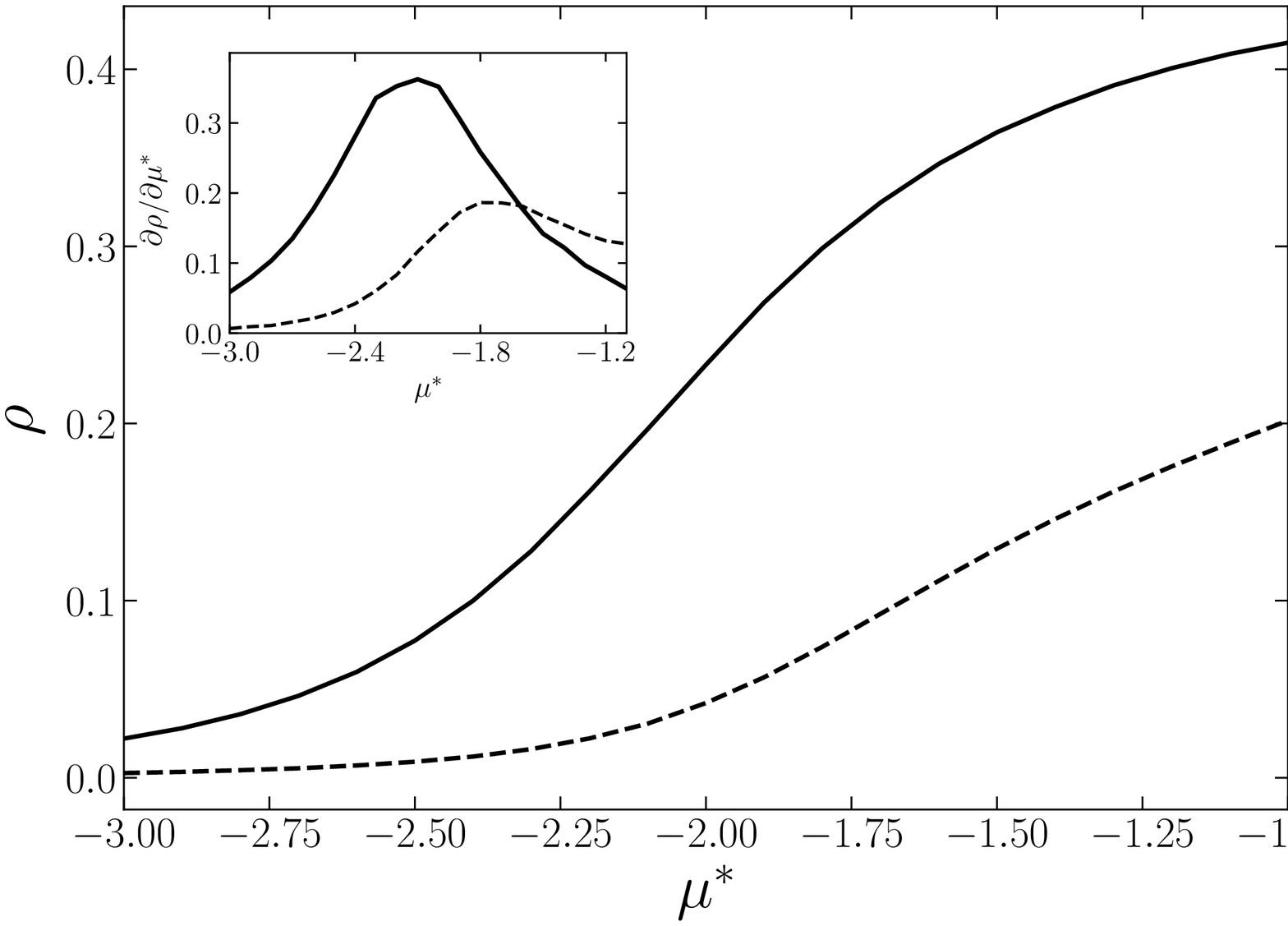}
	\caption{\label{fig:Densities} The average density in the row next to the wall
	(solid line) and in the bulk (dashed line) at $T^*=0.5$ and for wall-particle interaction $h = -1$.
	The derivative of the two densities with respect to the chemical potential is shown in the inset. 
	The turnover of the densities increase occurs at $\mu^*=-1.62$.}   
\end{figure}

Because of the attraction to the wall, the density of particles $\rho_0$ in the row adjoining to it grows 
initially faster with the chemical potential increase than in the bulk (Fig. ~\ref{fig:Densities}).
However, when $\rho_0$ attains the value corresponding to the maximal adsorption, its growth slows down.

The total adsorption is mainly determined by the rivalry of the density deviations in the two 
rows closest to the wall and the two subsequent ones that is the result of the competing interparticle interactions.
The long range repulsive interaction is of minor importance at low density of particles. With the chemical potential
increase and due to the wall attraction, the density of the particles in the layers closest to the wall increases
to the values when the interparticle repulsion joins the game and starts to hamper the density increase.
It is well illustrated by the dependence of the density $\rho_0$ on the intensity of the wall attraction 
(Fig. ~\ref{fig:profiles}). Even at the lowest bulk density (Fig. ~\ref{fig:profiles}a) the ratio ($\rho_0/\rho_c$) 
is considerably smaller than the Boltzmann factor exp$(-h/T^*)$, especially at $h=-2$. This is just the influence
of the interparticle repulsion, because $\rho_0$ is large enough when the repulsion can order the system into 
succession of rhombuses\cite{almarza:14:0}.  The upper bound for $\rho_0$ for a layer of clusters that do not 
repel one another is $\rho_0=1/2$ (see Fig.\ref{fig:GS_ADS}). At the same time, the bulk density increases as well, 
and the density in the rows 3 and 4  decreases to very small values. These two layers lead to a negative contribution
to the adsorption, and the absolute value of this contribution increases with increasing bulk density. The result of
this is the maximum on the chemical potential dependence of the total adsorption. 

An additional information can be drawn from adsorption at neutral and repulsive walls. The neutral wall behaves 
like an attractive one, especially at the chemical potentials corresponding to not very small concentrations
(Fig. ~\ref{fig:ADS}). Definitely, this is the result of the long range interparticle repulsion. 
The particles in the rows closest to the wall  feel repulsion from the bulk side of the system and do 
not experience such an action from the wall. This effect comes even out for the repulsive wall resulting 
in a considerably weaker negative adsorption as compared to that at the attractive one.   

\subsection{Cluster formation}

In systems with competing interactions the formation of clusters plays an important role in
general~\cite{almarza:14:0, bomont:12:0, santos:17:0} as well as in adsorption phenomena as it was demonstrated
for a 3D off-lattice system\cite{litniewski:19:0}. Thus, the analysis of possible clusters and their distribution
in the near wall region can give an additional insight into characterization of the adsorption in our 2D lattice system.  

\begin{figure}[htb!]
	\includegraphics[width=0.8\linewidth]{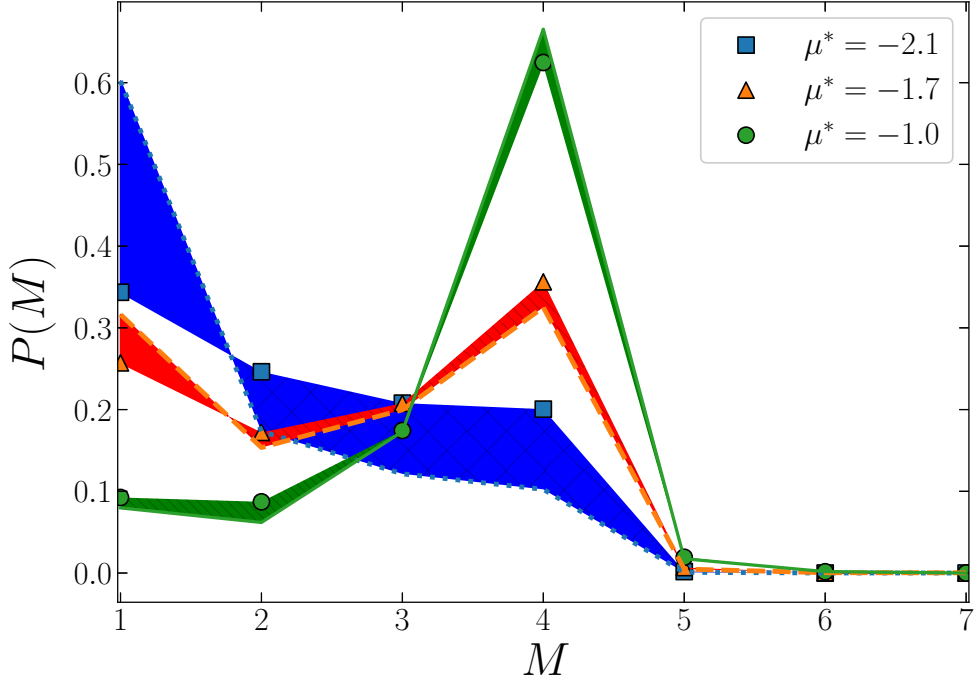}
	\caption{\label{fig:Cluster_distr} The distribution of the probabilities for particles to 
	belong to a cluster of size $M$ for different values of the chemical potential at $T^*=0.5$ in 
	the two closest to the wall rows and in the bulk. The partial contributions of different configurations 
	of particles in a cluster of size $M$ are not displayed. Symbols indicate the distributions in the 
	near-wall region (squares, triangles and circles are used for the chemical potentials $\mu^* = -2.1, -1.7, -1.0$, 
	respectively). The filled areas reflect the difference between the probability distribution for certain clusters 
	in the bulk and in the near-wall area. The hatched fillings indicate the probability excess in the border area as 
	compared to the bulk. For $\mu^*=-1.7$, $P(1)=0.316$ and $P(4)=0.325$ in the bulk. As these values are very close
	to each other, at this value of the chemical potential the monomer-dominated fluid crosses over to the 
	cluster-dominated fluid.}
\end{figure}

The following types of clusters are predominantly found in the gas phase of these systems: monomers, dimers,
2 types of triangular clusters and rhomboidal clusters. The contribution of each of them changes with increasing
the chemical potential or density both in the bulk and in the border area (Fig. ~\ref{fig:Cluster_distr}).

The distribution is shown for cluster sizes $M \leq 7$ since the contribution of clusters of higher 
sizes is much lower of the order of $10^{-6}$. In addition, for each of the sizes greater than or equal 
to 3, the sum of the probabilities of all possible configurations was calculated due to the fact that the 
probability of the most energetically favorable configuration is much higher than the others for the same 
cluster size (an equilateral triangle for $M$=3, a rhombus for $M$=4, a trapezoid for $M$=5) and is approximately
95\% relative to the total number of clusters of a given size for the chemical potential values considered.
As an exception, three-particle clusters can be distinguished: the contribution of the clusters that form 
the equilateral triangle is about 80\%, about 20\% is a part of the clusters, forming an irregular triangle, and 
much less than 1\% are for linear clusters. 

With the density increase, the particle distribution shifts to larger cluster sizes. As compared to the bulk, 
the excess of isolated particles in the near-wall region decreases with the chemical potential increase and 
takes a negative value for $\mu=-1.0$. At the chemical potential that corresponds to the maximum of the 
adsorption, the probability distribution of particles among the clusters is a qualitatively different from two 
other situations. The probability distribution has two highs, one for isolated particles and the other for rhombuses. 

A similar situation takes place in the bulk that can be explained by the density increase although there is a 
large difference in the bulk and the near-wall mean particle densities (0.2 against 0.4 at $\mu=-1$ and $T^*=0.5$).
The ordered structures in these regions are different. In the near-wall region a two-row stripe filled by the rhomboid
clusters is followed by almost empty two rows while in the bulk they are homogeneously distributed. Thus, the most 
ordered rhombus state in the bulk\cite{almarza:14:0} is observed at $\rho=1/3$, while in the near-wall region at 
$\rho=1/2$ (Fig. ~\ref{fig:GS_ADS}) that leads to similar distributions of particles among the clusters in both regions. 
Two orientations of rhomboid clusters in the near-wall region are observed with the edges parallel to the wall.

\subsection{Correlation effect}

For a detailed analysis, we decided to study the behavior of the correlation functions $g_z(\delta y)$ depending on
the distance $\delta y$ between the sites of the lattice in the direction parallel to the wall, for several layers 
at the distance $z$ from it.

\begin{equation}
    g_z(\delta y) = \langle \hat\rho(y,z)  \hat\rho(y+\delta y,z) \rangle 
\end{equation}

\begin{figure}[htb!]
	\includegraphics[width=1\linewidth]{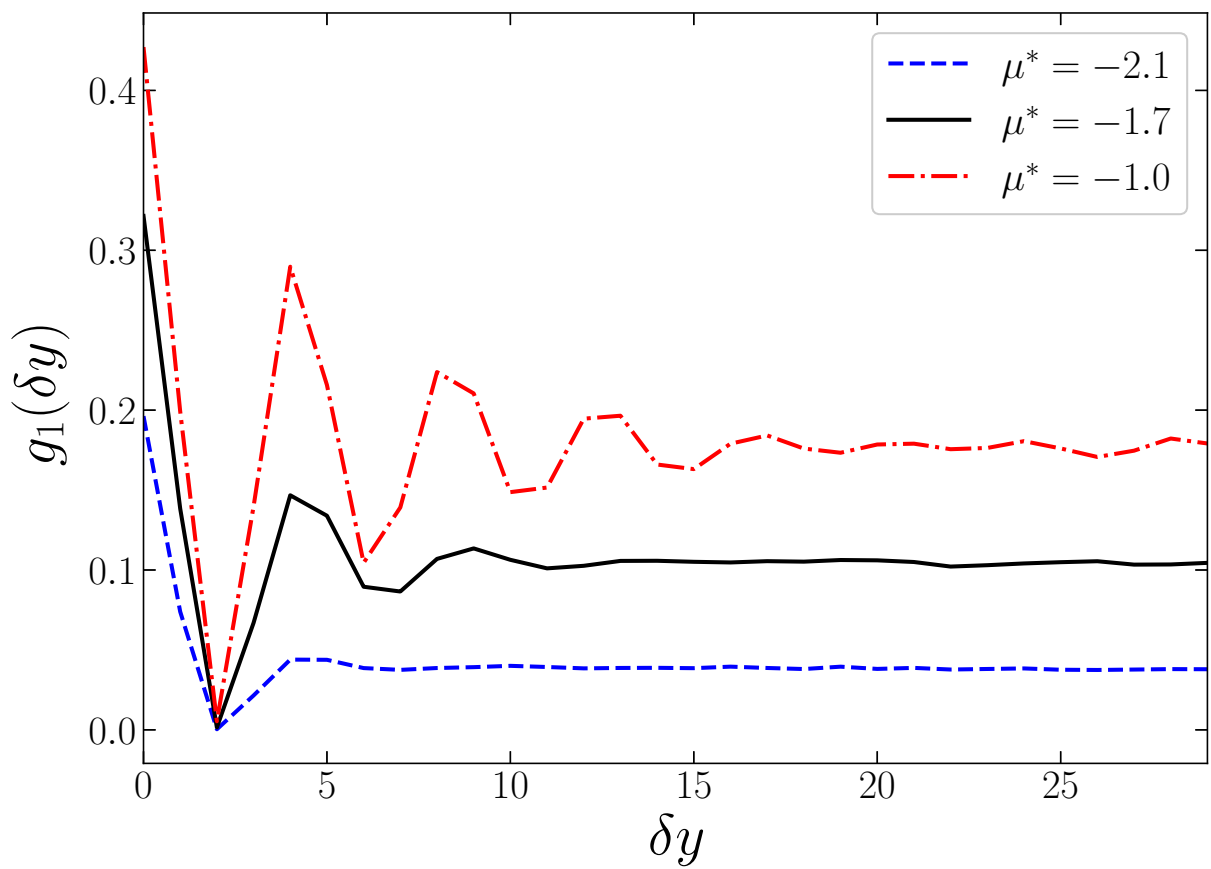}
	\caption{\label{fig:CorFunc} Correlation functions $g_1(\delta y)$ along the wall for the first row for 
	different values of the chemical potential, which correspond to the states before, at and after the maximum
	of the adsorption $\Gamma(\mu)$ at temperature $T^*=0.5$ and the particle-wall interaction energy $h=-1$.}
\end{figure}

For small values of $\mu^*$, when the density is very low, the density at the surface grows faster with increasing
the chemical potential than the density in the bulk, because interaction with the wall makes it favourable to 
introduce a particle or a cluster at the surface. When the density becomes larger, and the average distance between
the clusters at the surface is too small to introduce another cluster without causing a repulsion with the new and 
the existing clusters, it becomes more favourable to introduce a cluster to the bulk, where the density is smaller.
At this value of the chemical potential the correlation function (Fig. ~\ref{fig:CorFunc}) in the near-wall row starts 
to show an oscillatory decay. This short-range ordering of clusters allows to avoid repulsion between some pairs of
clusters if the distribution of clusters would be random. No correlation was observed in the positions of particles 
in the first and fifth rows due to very low density of particles in the third and fourth rows.

The oscillatory decay of the correlation function in the parallel direction and of the density profile in the 
perpendicular direction both start when the adsorption begins to decay as a function of $\mu^*$. This is 
consistent with the slower increase of the density at the wall than in the bulk. In the bulk the density 
is still low enough and the probability that a cluster introduced randomly will be close to existing clusters is low.

Since the density at the surface (rows 1 and 2) grows more slowly, and in the layer next to them (rows 3 and 4)
the density attains very small values, the difference between the mean near-wall density and the density in the 
bulk must decrease. This is an effect of the repulsion between the clusters when the density is relatively large.
It leads to the local periodic ordering on the one hand, and to the decreasing adsorption on the other hand. 

As a result of which, it can be noted that the attractive surface covered by clusters changes to an efficiently repulsive one: 
the repulsive barrier in it is formed by adsorbed particles\cite{litniewski:19:0}, which, due to interaction with each other,
have a stronger repulsion at long distances.

\section{\label{sec:concl}Summary and Conclusions}
The purpose of this work was investigation of the effect of cluster formation on adsorption phenomena. We focused on
a monolayer of SALR particles confined by a straight wall. In order to determine general effects common to many SALR systems,
we considered a generic model with a phase diagram determined earlier in Ref.\cite{almarza:14:0}. We assumed that the 
particles occupied sites of the 2D triangular lattice, and interacted as in the model introduced in Ref.~\cite{pekalski:14:0} 
(first neighbor attraction and third neighbor repulsion). The wall modeled by a straight line interacted only with the 
particles in the first row (next to the wall).

We have obtained the adsorption isotherm as a function of the chemical potential, $\Gamma(\mu^*)$, for a few values of
temperature and for different strengths of the wall-particle interaction. In addition, structural characteristics such
as cluster size distribution in the bulk and near the wall, density profile in the direction perpendicular to the wall,
correlation function in the direction parallel to the wall, density in the first row and in the bulk, and partial 
adsorption defined in Eq.(\ref{eq:PartAds}) were computed. All quantities were obtained by MC simulations.

We have found that the shape of the adsorption isotherm is qualitatively different than in simple fluids. In the case 
of the dilute phase, the adsorption is a nondecreasing function of the chemical potential when the long-range repulsion
between the particles is absent. In contrast, if the long-range repulsion is present, the adsorption takes a pronounced 
maximum for the chemical potential $\mu^*=\mu^*_{max}$ that is significantly smaller than its value at the phase transition
(i.e. still in the low-density disordered phase). We have found this characteristic non-standard shape of $\Gamma(\mu^*)$
for all studied temperatures and even in the absence of wall-particle attraction.

Interestingly, all the studied structural characteristics undergo a qualitative change for $\mu^*\approx\mu^*_{max}$.
When $\Gamma(\mu^*)$ is an increasing function of $\mu^*$, the system behavior is dominated by individual particles. 
Even though clusters are present when $\mu^*$ exceeds a certain value, the probability of finding an isolated particle 
is larger than the probability of finding a particle belonging to the optimal cluster (Fig.\ref{fig:Cluster_distr}).
The density in the first row increases faster than the density in the bulk for increasing $\mu^*$ (Fig.\ref{fig:Densities}), 
but it is still low enough so that the average distance between the particles is larger than the range of the repulsion. 
No short-range order is present near the wall. In this low-density regime, it is more probable that single particles rather
than clusters of particles will be introduced to the system upon increase of $\mu^*$. Moreover, with a large probability 
the new particles will be adsorbed at the attractive wall. Even a neutral wall effectively attracts particles, because
the long-range repulsion by the particles at $z>0$ is not compensated due to missing neighbors for $z<0$.

For $\mu^*>\mu^*_{max}$, however, the probability of finding an isolated particle is smaller than the probability of 
finding a particle belonging to the optimal cluster (Fig.\ref{fig:Cluster_distr}). In this case, we may expect that
clusters will be introduced to the system with a larger probability than isolated particles when $\mu^*$ increases. For 
this range of $\mu^*$, the density in the first row is larger, and the average distance between the clusters is significantly
smaller than in the bulk. For this reason the long-range repulsion can be more easily avoided when a new cluster is 
introduced to the bulk, rather than to the near-surface region. As a result, the density in the bulk grows faster than
at the wall, and  the adsorption decreases for increasing $\mu^*$. Moreover, to avoid the repulsion between the clusters,
a short-range order near the wall appears. This short-range order is represented by the oscillatory density and correlation
function in the perpendicular and parallel directions, respectively. The density in the layer of clusters at the wall 
(rows 1 and 2) approaches $1/2$, and the density in the rows 3 and 4 approaches $0$ when $T\to 0$ (Fig\ref{fig:Densities}).
This empty layer gives a negative contribution to the adsorption, and shows that an attractive surface covered by the
clusters of the SALR particles becomes effectively repulsive. A similar depletion zone was observed in a 3D 
system~\cite{litniewski:19:0}. We expect the remaining anomalies to hold in 3D as well. However, due to a 
repulsive barrier between the adsorbed particles and the bulk, it can be tricky to study and would require 
nontrivial sampling methods.

It would be interesting to verify our predictions experimentally. The maximum of $\Gamma(\mu^*)$ could serve as an 
indication of a crossover between monomer- and cluster-dominated fluid, and of appearance of short-range order 
near the system boundary.

\section{Acknowledgements}

This project has received funding from the European Union’s Horizon 2020 research and innovation program under 
the Marie Skłodowska-Curie grant agreement No 734276. Additional support in the years 2017–2020 has been granted 
for the CONIN project by the Polish Ministry of Science and Higher Education (agreement no. 3854/H2020/17/2018/2).
Financial support from the National Science Center under Grant No. 2015/19/B/ST3/03122 is also acknowledged.


\begin{thebibliography}{34}
\expandafter\ifx\csname natexlab\endcsname\relax\def\natexlab#1{#1}\fi
\expandafter\ifx\csname bibnamefont\endcsname\relax
  \def\bibnamefont#1{#1}\fi
\expandafter\ifx\csname bibfnamefont\endcsname\relax
  \def\bibfnamefont#1{#1}\fi
\expandafter\ifx\csname citenamefont\endcsname\relax
  \def\citenamefont#1{#1}\fi
\expandafter\ifx\csname url\endcsname\relax
  \def\url#1{\texttt{#1}}\fi
\expandafter\ifx\csname urlprefix\endcsname\relax\def\urlprefix{URL }\fi
\providecommand{\bibinfo}[2]{#2}
\providecommand{\eprint}[2][]{\url{#2}}

\bibitem[{\citenamefont{Roe}(1974)}]{roe:74:0}
\bibinfo{author}{\bibfnamefont{R.-J.} \bibnamefont{Roe}},
  \bibinfo{journal}{{\it J. Chem. Phys.}} \textbf{\bibinfo{volume}{60}},
  \bibinfo{pages}{4192} (\bibinfo{year}{1974}).

\bibitem[{\citenamefont{Dash}(1975)}]{dash:75:0}
\bibinfo{author}{\bibfnamefont{J.~G.} \bibnamefont{Dash}},
  \emph{\bibinfo{title}{Films on solid surfaces: the physics and chemistry of
  physical adsorption}} (\bibinfo{publisher}{Elsevier}, \bibinfo{year}{1975}).

\bibitem[{\citenamefont{Kreuzer and Gortel}(1986)}]{kreuzer:86:0}
\bibinfo{author}{\bibfnamefont{H.}~\bibnamefont{Kreuzer}} \bibnamefont{and}
  \bibinfo{author}{\bibfnamefont{Z.}~\bibnamefont{Gortel}},
  \textbf{\bibinfo{volume}{1}} (\bibinfo{year}{1986}).

\bibitem[{\citenamefont{Kukushkin and Osipov}(1998)}]{kukushkin:98:0}
\bibinfo{author}{\bibfnamefont{S.~A.} \bibnamefont{Kukushkin}}
  \bibnamefont{and} \bibinfo{author}{\bibfnamefont{A.~V.}
  \bibnamefont{Osipov}}, \bibinfo{journal}{{\it Physics-Uspekhi}}
  \textbf{\bibinfo{volume}{41}}, \bibinfo{pages}{983} (\bibinfo{year}{1998}).

\bibitem[{\citenamefont{Jerome}(1991)}]{jerome:91:0}
\bibinfo{author}{\bibfnamefont{B.}~\bibnamefont{Jerome}},
  \bibinfo{journal}{{\it Reports on Progress in Physics}}
  \textbf{\bibinfo{volume}{54}}, \bibinfo{pages}{391} (\bibinfo{year}{1991}).

\bibitem[{\citenamefont{Patrykiejew et~al.}(2000)\citenamefont{Patrykiejew,
  Soko{\l}owski, and Binder}}]{patrykiejew:00:0}
\bibinfo{author}{\bibfnamefont{A.}~\bibnamefont{Patrykiejew}},
  \bibinfo{author}{\bibfnamefont{S.}~\bibnamefont{Soko{\l}owski}},
  \bibnamefont{and} \bibinfo{author}{\bibfnamefont{K.}~\bibnamefont{Binder}},
  \bibinfo{journal}{{ \it Surface science reports}}
  \textbf{\bibinfo{volume}{37}}, \bibinfo{pages}{207} (\bibinfo{year}{2000}).

\bibitem[{\citenamefont{Netz and Andelman}(2003)}]{netz:03:0}
\bibinfo{author}{\bibfnamefont{R.~R.} \bibnamefont{Netz}} \bibnamefont{and}
  \bibinfo{author}{\bibfnamefont{D.}~\bibnamefont{Andelman}},
  \bibinfo{journal}{{\it Physics reports}} \textbf{\bibinfo{volume}{380}},
  \bibinfo{pages}{1} (\bibinfo{year}{2003}).

\bibitem[{\citenamefont{Bruch et~al.}(2007)\citenamefont{Bruch, Cole, and
  Zaremba}}]{bruch:07:0}
\bibinfo{author}{\bibfnamefont{L.~W.} \bibnamefont{Bruch}},
  \bibinfo{author}{\bibfnamefont{M.~W.} \bibnamefont{Cole}}, \bibnamefont{and}
  \bibinfo{author}{\bibfnamefont{E.}~\bibnamefont{Zaremba}},
  \emph{\bibinfo{title}{Physical adsorption: forces and phenomena}}
  (\bibinfo{publisher}{Courier Dover Publications}, \bibinfo{year}{2007}).

\bibitem[{\citenamefont{Rabe et~al.}(2011)\citenamefont{Rabe, Verdes, and
  Seeger}}]{rabe:11:0}
\bibinfo{author}{\bibfnamefont{M.}~\bibnamefont{Rabe}},
  \bibinfo{author}{\bibfnamefont{D.}~\bibnamefont{Verdes}}, \bibnamefont{and}
  \bibinfo{author}{\bibfnamefont{S.}~\bibnamefont{Seeger}},
  \bibinfo{journal}{{\it Advances in colloid and interface science}}
  \textbf{\bibinfo{volume}{162}}, \bibinfo{pages}{87} (\bibinfo{year}{2011}).

\bibitem[{\citenamefont{Antelmi et~al.}(1995)\citenamefont{Antelmi,
  K\'ekicheff, and Richetti}}]{antelmi:95:0}
\bibinfo{author}{\bibfnamefont{D.}~\bibnamefont{Antelmi}},
  \bibinfo{author}{\bibfnamefont{P.}~\bibnamefont{K\'ekicheff}},
  \bibnamefont{and} \bibinfo{author}{\bibfnamefont{P.}~\bibnamefont{Richetti}},
  \bibinfo{journal}{{\it J. Phys. II France}} \textbf{\bibinfo{volume}{5}},
  \bibinfo{pages}{103} (\bibinfo{year}{1995}).

\bibitem[{\citenamefont{Yu et~al.}(2006)\citenamefont{Yu, Sun, Chen, Jin, Ding,
  Li, and Shi}}]{yu:06:0}
\bibinfo{author}{\bibfnamefont{B.}~\bibnamefont{Yu}},
  \bibinfo{author}{\bibfnamefont{P.}~\bibnamefont{Sun}},
  \bibinfo{author}{\bibfnamefont{T.}~\bibnamefont{Chen}},
  \bibinfo{author}{\bibfnamefont{Q.}~\bibnamefont{Jin}},
  \bibinfo{author}{\bibfnamefont{D.}~\bibnamefont{Ding}},
  \bibinfo{author}{\bibfnamefont{B.}~\bibnamefont{Li}}, \bibnamefont{and}
  \bibinfo{author}{\bibfnamefont{A.-C.} \bibnamefont{Shi}},
  \bibinfo{journal}{{\it Phys. Rev. Lett.}} \textbf{\bibinfo{volume}{96}},
  \bibinfo{pages}{138306} (\bibinfo{year}{2006}).

\bibitem[{\citenamefont{Yu et~al.}(2008)\citenamefont{Yu, Jin, Ding, Li, and
  Shi}}]{yu:08:0}
\bibinfo{author}{\bibfnamefont{B.}~\bibnamefont{Yu}},
  \bibinfo{author}{\bibfnamefont{Q.}~\bibnamefont{Jin}},
  \bibinfo{author}{\bibfnamefont{D.}~\bibnamefont{Ding}},
  \bibinfo{author}{\bibfnamefont{B.}~\bibnamefont{Li}}, \bibnamefont{and}
  \bibinfo{author}{\bibfnamefont{A.-C.} \bibnamefont{Shi}},
  \bibinfo{journal}{{\it Macromolecules}} \textbf{\bibinfo{volume}{41}},
  \bibinfo{pages}{4042} (\bibinfo{year}{2008}).

\bibitem[{\citenamefont{Su et~al.}(2003)\citenamefont{Su, Guo, and
  Palmer}}]{su:03:0}
\bibinfo{author}{\bibfnamefont{G.}~\bibnamefont{Su}},
  \bibinfo{author}{\bibfnamefont{Q.}~\bibnamefont{Guo}}, \bibnamefont{and}
  \bibinfo{author}{\bibfnamefont{R.}~\bibnamefont{Palmer}},
  \bibinfo{journal}{{\it Langmuir}} \textbf{\bibinfo{volume}{19}},
  \bibinfo{pages}{9669} (\bibinfo{year}{2003}).

\bibitem[{\citenamefont{Li et~al.}(2005)\citenamefont{Li, Xing, Huang, and
  Han}}]{li:05:0}
\bibinfo{author}{\bibfnamefont{J.}~\bibnamefont{Li}},
  \bibinfo{author}{\bibfnamefont{R.}~\bibnamefont{Xing}},
  \bibinfo{author}{\bibfnamefont{W.}~\bibnamefont{Huang}}, \bibnamefont{and}
  \bibinfo{author}{\bibfnamefont{Y.}~\bibnamefont{Han}}, \bibinfo{journal}{{\it
  Colloids and Surfaces A: Physicochemical and Engineering Aspects}}
  \textbf{\bibinfo{volume}{269}}, \bibinfo{pages}{22} (\bibinfo{year}{2005}).

\bibitem[{\citenamefont{Huang et~al.}(2006)\citenamefont{Huang, Li, Luo, Zhang,
  Luan, and Han}}]{huang:06:0}
\bibinfo{author}{\bibfnamefont{W.}~\bibnamefont{Huang}},
  \bibinfo{author}{\bibfnamefont{J.}~\bibnamefont{Li}},
  \bibinfo{author}{\bibfnamefont{C.}~\bibnamefont{Luo}},
  \bibinfo{author}{\bibfnamefont{J.}~\bibnamefont{Zhang}},
  \bibinfo{author}{\bibfnamefont{S.}~\bibnamefont{Luan}}, \bibnamefont{and}
  \bibinfo{author}{\bibfnamefont{Y.}~\bibnamefont{Han}}, \bibinfo{journal}{{\it
  Colloids and Surfaces }} \textbf{\bibinfo{volume}{273}}, \bibinfo{pages}{43}
  (\bibinfo{year}{2006}).

\bibitem[{\citenamefont{Haghgooie et~al.}(2006)\citenamefont{Haghgooie, Li, and
  Doyle}}]{haghgooie:06:0}
\bibinfo{author}{\bibfnamefont{R.}~\bibnamefont{Haghgooie}},
  \bibinfo{author}{\bibfnamefont{C.}~\bibnamefont{Li}}, \bibnamefont{and}
  \bibinfo{author}{\bibfnamefont{P.~S.} \bibnamefont{Doyle}},
  \bibinfo{journal}{{\it Langmuir}} \textbf{\bibinfo{volume}{22}},
  \bibinfo{pages}{3601} (\bibinfo{year}{2006}).

\bibitem[{\citenamefont{Tasinkevych and Ciach}(2001)}]{tasinkevych:01:0}
\bibinfo{author}{\bibfnamefont{M.}~\bibnamefont{Tasinkevych}} \bibnamefont{and}
  \bibinfo{author}{\bibfnamefont{A.}~\bibnamefont{Ciach}},
  \bibinfo{journal}{{\it J. Chem. Phys.}} \textbf{\bibinfo{volume}{115}},
  \bibinfo{pages}{8705} (\bibinfo{year}{2001}).

\bibitem[{\citenamefont{Imperio and Reatto}(2007)}]{imperio:07:0}
\bibinfo{author}{\bibfnamefont{A.}~\bibnamefont{Imperio}} \bibnamefont{and}
  \bibinfo{author}{\bibfnamefont{L.}~\bibnamefont{Reatto}},
  \bibinfo{journal}{{\it Phys. Rev. E}} \textbf{\bibinfo{volume}{76}},
  \bibinfo{pages}{040402} (\bibinfo{year}{2007}).

\bibitem[{\citenamefont{Imperio et~al.}(2008)\citenamefont{Imperio, Reatto, and
  Zapperi}}]{imperio:08:0}
\bibinfo{author}{\bibfnamefont{A.}~\bibnamefont{Imperio}},
  \bibinfo{author}{\bibfnamefont{L.}~\bibnamefont{Reatto}}, \bibnamefont{and}
  \bibinfo{author}{\bibfnamefont{S.}~\bibnamefont{Zapperi}},
  \bibinfo{journal}{{\it Phys. Rev. E}} \textbf{\bibinfo{volume}{78}},
  \bibinfo{pages}{021402} (\bibinfo{year}{2008}).


\bibitem[{\citenamefont{Archer}(2008)}]{archer:08:0}
\bibinfo{author}{\bibfnamefont{A.~J.} \bibnamefont{Archer}},
  \bibinfo{journal}{{\it Phys. Rev. E}} \textbf{\bibinfo{volume}{78}},
  \bibinfo{pages}{031402} (\bibinfo{year}{2008}).

\bibitem[{\citenamefont{Chi et~al.}(2011)\citenamefont{Chi, Wang, Li, and
  Shi}}]{chi:11:0}
\bibinfo{author}{\bibfnamefont{P.}~\bibnamefont{Chi}},
  \bibinfo{author}{\bibfnamefont{Z.}~\bibnamefont{Wang}},
  \bibinfo{author}{\bibfnamefont{B.}~\bibnamefont{Li}}, \bibnamefont{and}
  \bibinfo{author}{\bibfnamefont{A.-C.} \bibnamefont{Shi}},
  \bibinfo{journal}{{\it Langmuir}} \textbf{\bibinfo{volume}{27}},
  \bibinfo{pages}{11683} (\bibinfo{year}{2011}).

\bibitem[{\citenamefont{Almarza et~al.}(2016)\citenamefont{Almarza,
  P\c{e}kalski, and Ciach}}]{almarza:16:0}
\bibinfo{author}{\bibfnamefont{N.~G.} \bibnamefont{Almarza}},
  \bibinfo{author}{\bibfnamefont{J.}~\bibnamefont{P\c{e}kalski}},
  \bibnamefont{and} \bibinfo{author}{\bibfnamefont{A.}~\bibnamefont{Ciach}},
  \bibinfo{journal}{\textit{Soft Matter}} \textbf{\bibinfo{volume}{12}},
  \bibinfo{pages}{7551} (\bibinfo{year}{2016}).

\bibitem[{\citenamefont{P{\c e}kalski et~al.}(2019)\citenamefont{P{\c e}kalski,
  Bildanau, and Ciach}}]{pekalski:19:0}
\bibinfo{author}{\bibfnamefont{J.}~\bibnamefont{P{\c e}kalski}},
  \bibinfo{author}{\bibfnamefont{E.}~\bibnamefont{Bildanau}}, \bibnamefont{and}
  \bibinfo{author}{\bibfnamefont{A.}~\bibnamefont{Ciach}},
  \bibinfo{journal}{{\it Soft Matter}} pp.~\bibinfo{pages}{--}
  (\bibinfo{year}{2019}), \urlprefix\url{http://dx.doi.org/10.1039/C9SM01179J}.

\bibitem[{\citenamefont{P\c{e}kalski et~al.}(2014)\citenamefont{P\c{e}kalski,
  Rogowski, and Ciach}}]{pekalski:14:1}
\bibinfo{author}{\bibfnamefont{J.}~\bibnamefont{P\c{e}kalski}},
  \bibinfo{author}{\bibfnamefont{P.}~\bibnamefont{Rogowski}}, \bibnamefont{and}
  \bibinfo{author}{\bibfnamefont{A.}~\bibnamefont{Ciach}},
  \bibinfo{journal}{{\it Mol. Phys}} \textbf{\bibinfo{volume}{113}},
  \bibinfo{pages}{1022} (\bibinfo{year}{2014}).

\bibitem[{\citenamefont{P{\c e}kalski et~al.}(2014)\citenamefont{P{\c e}kalski,
  Ciach., and Almarza}}]{pekalski:14:0}
\bibinfo{author}{\bibfnamefont{J.}~\bibnamefont{P{\c e}kalski}},
  \bibinfo{author}{\bibfnamefont{A.}~\bibnamefont{Ciach.}}, \bibnamefont{and}
  \bibinfo{author}{\bibfnamefont{N.~G.} \bibnamefont{Almarza}},
  \bibinfo{journal}{{\it J. Chem. Phys.}} \textbf{\bibinfo{volume}{140}},
  \bibinfo{pages}{114701} (\bibinfo{year}{2014}).

\bibitem[{\citenamefont{Santos et~al.}(2017)\citenamefont{Santos, P\c{e}kalski,
  and Panagiotopoulos}}]{santos:17:0}
\bibinfo{author}{\bibfnamefont{A.~P.} \bibnamefont{Santos}},
  \bibinfo{author}{\bibfnamefont{J.}~\bibnamefont{P\c{e}kalski}},
  \bibnamefont{and} \bibinfo{author}{\bibfnamefont{A.~Z.}
  \bibnamefont{Panagiotopoulos}}, \bibinfo{journal}{{\it Soft Matter}}
  \textbf{\bibinfo{volume}{13}}, \bibinfo{pages}{8055} (\bibinfo{year}{2017}).

\bibitem[{\citenamefont{Litniewski and Ciach}(2019)}]{litniewski:19:0}
\bibinfo{author}{\bibfnamefont{M.}~\bibnamefont{Litniewski}} \bibnamefont{and}
  \bibinfo{author}{\bibfnamefont{A.}~\bibnamefont{Ciach}},
  \bibinfo{journal}{{\it J. Chem. Phys.}} \textbf{\bibinfo{volume}{150}},
  \bibinfo{pages}{234702} (\bibinfo{year}{2019}).

\bibitem[{\citenamefont{Zhuang et~al.}(2016)\citenamefont{Zhuang, Zhang, and
  Charbonneau}}]{zhuang:16:0}
\bibinfo{author}{\bibfnamefont{Y.}~\bibnamefont{Zhuang}},
  \bibinfo{author}{\bibfnamefont{K.}~\bibnamefont{Zhang}}, \bibnamefont{and}
  \bibinfo{author}{\bibfnamefont{P.}~\bibnamefont{Charbonneau}},
  \bibinfo{journal}{{\it Phys. Rev. Lett.}} \textbf{\bibinfo{volume}{116}},
  \bibinfo{pages}{098301} (\bibinfo{year}{2016}).

\bibitem[{\citenamefont{Royall}(2018)}]{royall:18:0}
\bibinfo{author}{\bibfnamefont{P.}~\bibnamefont{Royall}},
  \bibinfo{journal}{{\it Soft matter}}  (\bibinfo{year}{2018}).

\bibitem[{\citenamefont{Zhuang and Charbonneau}(2016)}]{zhuang:16:2}
\bibinfo{author}{\bibfnamefont{Y.}~\bibnamefont{Zhuang}} \bibnamefont{and}
  \bibinfo{author}{\bibfnamefont{P.}~\bibnamefont{Charbonneau}},
  \bibinfo{journal}{{\it J. Phys. Chem. B}} \textbf{\bibinfo{volume}{120}},
  \bibinfo{pages}{7775} (\bibinfo{year}{2016}).

\bibitem[{\citenamefont{Stradner et~al.}(2004)\citenamefont{Stradner, Sedgwick,
  Cardinaux, Poon, Egelhaaf, and Schurtenberger}}]{stradner:04:0}
\bibinfo{author}{\bibfnamefont{A.}~\bibnamefont{Stradner}},
  \bibinfo{author}{\bibfnamefont{H.}~\bibnamefont{Sedgwick}},
  \bibinfo{author}{\bibfnamefont{F.}~\bibnamefont{Cardinaux}},
  \bibinfo{author}{\bibfnamefont{W.}~\bibnamefont{Poon}},
  \bibinfo{author}{\bibfnamefont{S.}~\bibnamefont{Egelhaaf}}, \bibnamefont{and}
  \bibinfo{author}{\bibfnamefont{P.}~\bibnamefont{Schurtenberger}},
  \bibinfo{journal}{{\it Nature}} \textbf{\bibinfo{volume}{432}},
  \bibinfo{pages}{492} (\bibinfo{year}{2004}).

\bibitem[{\citenamefont{Almarza et~al.}(2014)\citenamefont{Almarza, P{\c
  e}kalski, and Ciach}}]{almarza:14:0}
\bibinfo{author}{\bibfnamefont{N.~G.} \bibnamefont{Almarza}},
  \bibinfo{author}{\bibfnamefont{J.}~\bibnamefont{P{\c e}kalski}},
  \bibnamefont{and} \bibinfo{author}{\bibfnamefont{A.}~\bibnamefont{Ciach}},
  \bibinfo{journal}{{\it J. Chem. Phys.}} \textbf{\bibinfo{volume}{140}},
  \bibinfo{pages}{164708} (\bibinfo{year}{2014}).

\bibitem[{\citenamefont{Bomont and Costa}(2012)}]{bomont:12:0}
\bibinfo{author}{\bibfnamefont{J.-M.} \bibnamefont{Bomont}} \bibnamefont{and}
  \bibinfo{author}{\bibfnamefont{D.}~\bibnamefont{Costa}},
  \bibinfo{journal}{{\it J. Chem. Phys.}} \textbf{\bibinfo{volume}{137}},
  \bibinfo{pages}{164901} (\bibinfo{year}{2012}).

\end{thebibliography}
\end{document}